# Automated Guidance of Collective Movement in a Multi-Agent Model of *Physarum polycephalum*

Jeff Jones

Centre for Unconventional Computing, University of the West of England, Bristol, UK
(E-mail: jeff.jones@uwe.ac.uk)

**Abstract:** Collective movement occurs in living systems where the simple movements of individual members of a population are combined to generate movement of the collective as a whole, displaying complex dynamics which cannot be found in the component parts themselves. The plasmodium stage of slime mould *Physarum polycephalum* displays complex amoeboid movement during its foraging and hazard avoidance and its movement can be influenced by the spatial placement of attractant and repellent stimuli. Slime mould is attractive to robotics due to its simple component parts and the distributed nature of its control and locomotion mechanisms. We investigate methods of automated guidance of a multi-agent swarm collective along a pre-defined path to a goal location. We demonstrate a closed-loop feedback mechanism using attractant and repellent stimuli. We find that guidance by repellent stimuli (a light illumination mask) provides faster and more accurate guidance than attractant sources, which exhibit overshooting phenomena at path turns. The method allows traversal of convoluted arenas with challenging obstacles and provides an insight into how unconventional computing substrates may be hybridised with classical computing methods to take advantage of the benefits of both approaches.

## 1. COLLECTIVE MOVEMENT

Collective movement is a directed movement of multiple individuals which are coupled (directly or indirectly) by some aspect of their environment or special senses. The phenomenon is observed in natural systems which span huge variations in spatial scale, temporal scale, and in their environmental medium. Collective movement at the population level can result in dynamic and dramatic patterning phenomena, such as swarming [6, 9], flocking [24], herding [8], and shoaling and schooling [13], [23]. The specific biological and generalised coupling mechanisms which generate these emergent patterns from the low-level individual interactions have been studied [27], [7], [34]. In human environments collective movement is seen in walking trail patterns [12], crowd dynamics [38], and car traffic systems [21].

Collective movement is of interest for robotics applications because the mechanisms which enable collective movement in natural systems are relatively simple, use local communication cues, exploit self-organised patterning and exhibit emergent behaviours and distributed control. This results in efficient collectives which contain redundant parts and are resilient to damage or interruption. Simple identical components would reduce the cost of robotic devices. Furthermore, communication between robotic entities and evaluation of current position and future goal position imply a significant computational cost which would be multiplied when the collective comprises a large robotic population. The use of strategies exploited by living collective systems in artificial robotic collectives may result in useful physical and computational cost savings, with the benefit of innate autonomy and resilience.

## 2. COLLECTIVE MOVEMENT IN SLIME MOULD

Slime mould *Physarum polycephalum* is a giant single-celled amoeboid organism, visible to the naked eye. During the plasmodium stage of its complex life-cycle [26] it takes the form of a constantly adapting protoplasmic network. This network is comprised of a sponge-like material which exhibits self-organised contractile oscillatory activity which is harnessed in the transport and distribution of nutrients within this internal transport network. The organism is remarkable in that the control of the oscillatory behaviour is distributed throughout the almost homogeneous medium and is highly redundant, having no critical or unique components.

The *P. polycephalum* plasmodium can thus be regarded as a complex functional material capable of both sensory and motor behaviour, in reference to the rich computational potential afforded by its material properties [16]. A degree of control over *P. polycephalum* is possible by exploiting its response to changing conditions within its environment. The migration of the plasmodium is influenced by a wide number of external stimuli including chemoattractants and chemorepellents [35], light irradiation [22], thermal gradients [36], substrate hardness [29], tactile stimulation [3], geotaxis [37] and magnetotaxis [25]. By careful manipulation of such external stimuli the plasmodium may be considered as a prototype robotic micro-mechanical manipulation system, capable of simple and programmable robotic actions including the manipulation (pushing and pulling) of small scale objects [5], [31], transport and mixing of substances [2].

*Physarum* has been shown to be a useful model organism in the study of distributed robotics. In this article we explore the problem of collective guidance, i.e. how to move and guide a population of independent mobile entities along a pre-determined path. This task represents a only small subset of general robotics challenges which also include the problem of how to survey and map an unknown environment, and how to plan paths between two or more locations in an environment. Approaches to robotics guidance and planning problems directly inspired by *Physarum* include the simultaneous localisation and mapping problem [20], the generation and manual guidance of collective transport [19], and amoeboid movement [18]. Mechanical instances of *Physarum*-inspired robots were demonstrated in [33].

---

† Jeff Jones is the presenter of this paper.

Hybrid natural computing and classical computing approaches are relatively uncommon. In the work of [4] a path planning system was implemented by using waves from a chemical reaction-diffusion processor to represent start points, end points and obstacles. These waves were used to generate a repulsive field which was used to guide a robot along the arena. The *Physarum* plasmodium itself was used as a guidance mechanism in a biological-mechanical hybrid approach where the response of the plasmodium to light irradiation stimuli provided by extended sensors from a classical robot device was then used to provide feedback control to the robot's movement actuators [32]. More recently the problem of generating a path between two points in an arena was tackled with a *Physarum*-inspired morphological adaptation approach [17].

We take the next logical step in these robotics challenges by tackling the problem of dynamically guiding a collective of mobile entities along the path whilst avoiding obstacles. We give an overview of the multi-agent modelling approach in Section 3. A hybrid approach to robotic guidance is presented in Section 4 with assessments of both attractant and repellent guidance methods, and some novel properties seen during path traversal. We conclude in Section 5 with a summary of the approach, its main properties and contribution.

## 3. MODELLING APPROACH

Multi-agent modelling has been an alternative to numerical models of *P. polycephalum* because they exhibit similar phenomena: simple components, local interactions, and emergent behaviour [10, 11]. In this report we use multi-agent approach introduced in [15] which generated dynamical adaptive transport networks. The approach has shown to be successful in reproducing a wide range of behaviour seen in *P. polycephalum*. Presentation of external environmental stimuli (both attractant and repellent) has been shown to be a critical factor in the evolution of patterning and complexity of computational behaviour within this model (for more information, see [16]). In this model the plasmodium is represented by a population of mobile particles with very simple behaviours, within a 2D diffusive environment. A discrete 2D lattice stores particle positions and also the concentration of a local generic chemoattractant. The chemoattractant concentration represents the hypothetical flux of sol within the plasmodium. Free particle movement represents the sol phase of the plasmodium and particle positions represent the fixed gel structure (i.e. global pattern) of the plasmodium. Particles act independently and their behaviour is divided into two distinct stages, the sensory stage and the motor stage. In the sensory stage, the particles sample their local environment using three forward biased sensors whose angle from the forward position (the sensor angle parameter, $SA$, set to 90°), and distance (sensor offset, $SO$, set to 15 pixels) may be parametrically adjusted. The offset sensors represent the overlapping filaments within the plasmodium, generating local indirect coupling of sensory inputs and movement to form networks of particles. The $SO$ distance is measured in pixels and the coupling effect increases as $SO$ increases.

During the sensory stage each particle changes its orientation to rotate (via the parameter rotation angle, $RA$, set to 45°) towards the strongest local source of chemoattractant. After the sensory stage, each particle executes the motor stage and attempts to move forwards in its current orientation (an angle from 0 - 360°) by a single pixel. Each lattice site may only store a single particle and particles deposit chemoattractant (5 arbitrary units) into the lattice only in the event of a successful forwards movement. If the next chosen site is already occupied by another particle the default (non-oscillatory) motor behaviour is to abandon the move, remain in the current position, and select a new random direction.

Diffusion of the collective chemoattractant signal is achieved via a simple $3 \times 3$ mean filter kernel with the mean multiplied by a damping parameter (set to 0.1) to limit the diffusion distance of the chemoattractant. The low level particle interactions result in complex pattern formation. The population spontaneously forms dynamic transport networks showing complex evolution and quasi-physical emergent properties, including closure of network lacunae, apparent surface tension effects and network minimisation. An exploration of the possible patterning parameters was presented in [14].

## 4. GUIDED COLLECTIVE MOVEMENT

To generate guided collective movement we utilise the self-oscillatory behaviour of the model plasmodium. This behaviour was introduced in [30] to reproduce the spontaneous and self-organised oscillation patterns observed within small samples of *Physarum* plasmodia [28]. In the model plasmodium these oscillations emerge from interruptions of individual particle movement. In the default non-oscillatory conditions a particle which is obstructed randomly selects a new orientation. In the oscillatory condition, however, the persistence of direction continues (by incrementing an internal position, compared to the current position of the particle) until a vacant space occurs along the current orientation path of the particle. It is possible to adjust the persistence time using a parameter $pID$ which tests for the probability of restoring the internal positional reference of the particle to its current actual position. Higher values of $pID$ (for exam-ple 0.05) results in less persistence of direction of the blob, whereas lower values (for example 0.001) results in much stronger persistence of direction. This parameter may therefore be used to control the momentum of the collective. The accumulation of interruptions in movement of individual particles results in travelling waves of flux forming within the mass of particles as particles occupy vacant spaces within the collective. It was shown in [18] that these travelling waves could shift the mass of particles, effectively moving the blob of virtual plasmodium. In the same paper it was demonstrated how the self-oscillatory dynamics could be influenced by the manual placement of attractant stimuli and simulated light irradiation (repellent) stimuli, causing the blobs to move towards attractants and away from light hazards.

### 4.1. Hybrid Control System

An oscillating blob of particles exhibits random movement when external stimuli are not present. The randomness of movement is due to the stochastic influences on particle orientation. This unpredictability of movement renders it challenging to control and guide their movement. A method for automatic guidance must represent a hybrid approach between the unconventional computing methods which generate the emergent behaviour in the virtual plasmodium (the generation of self-oscillatory travelling waves and amoeboid movement), and classical computing methods to detect the position of the blob and provide the feedback stimuli to guide the blob along the chosen path. A schematic overview of the closed-loop hybrid system is given in Fig. 1.

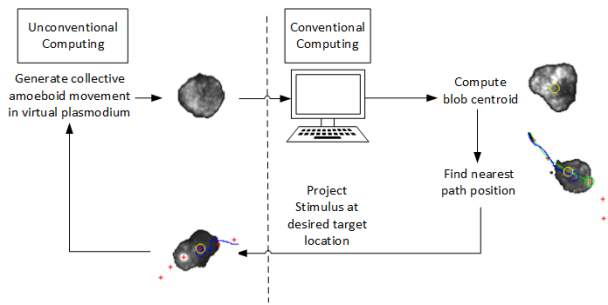

Fig. 1 Schematic overview of closed-loop guidance method. Left side of dashed line represents the contribution of the swarm approach, characterised by self-organised emergent phenomena, right side of line represents the contribution of the control method.

Note that Fig. 1 is partitioned by a vertical dashed line. This line indicates the separation of conventional and unconventional approaches and also indicates regions where both approaches interact. The unconventional part of the system generates the emergent oscillatory behaviour of the blob from local and self-organised particle interactions. Information about the blob's collective state is then extracted by the conventional (classical) part of the method which calculates the centroid (centre of mass) of the blob. The position of the blob is compared at every 50 scheduler steps to the points comprising the path in the arena. When the blob is closer to the next point along the path than to the current stimulus location, the *next* point along the path is then selected to provide the new location stimulus for the blob. The target stimulus is then projected into the spatially implemented unconventional part of the method. This stimulus acts to guide the blob towards this new location. Two possible stimulus types can be used, both of which are seen in the *Physarum* plasmodium and are described in the following sections.

### 4.2. Automatic Guidance with Attractant Stimuli

The first stimulus method used attractant stimuli. In the wild, slime mould is attracted to certain stimuli such as nutrients or localised warm areas, streaming towards these regions. In the model plasmodium attractant stimuli are represented by projecting localised stimuli into the diffusive lattice. These stimuli, when projected within sensory range of the blob, attract the particles comprising the blob. Particles at the outer periphery of the blob move towards the stimulus and the cohesion of the blob (caused by the indirect coupling of the particles' offset sensors) generates travelling waves inside the blob which move in the direction of the stimulus. As demonstrated by manual placement in [18], this can be used to guide (or 'pull') the blob *towards* the chosen direction. In the closed-loop method described in this article we can replace the manual placement of stimuli with automated transient placement of point stimuli to guide the blob along the chosen path.

Fig. 2 shows the results of automated closed-loop guidance of the self-oscillatory blob along a pre-defined path by the attractant method. The path starts at the large cross marker and ends at the circle marker, and individual guidance points on the path are denoted by small crosses (Fig. 2a). The path is composed of multiple links between start and end points in an arena populated by solid obstacles (grey shapes) which the agent particles cannot cross. The particle population, comprising 2000 particles, was inoculated at the initial cross marker position and for the first 1000 scheduler steps the blob was allowed to form and stabilise. After 1000 steps the automated guidance mechanism described in the previous sub-section was initiated. The short straight lines connecting the small crosses indicates the pre-defined path (green, online) and the path taken by the oscillatory blob is indicated by the blue (online) markers. The example results include two different momentum ($pID$) parameter settings.

Although the blob follows the path in all examples, at low $pID$ (i.e. high momentum) settings there is considerable deviation from the desired path, particularly when the path changes direction (see, for example, Fig. 2a).

This overshooting of the desired path is caused by the blob position being influenced by the strong oscillatory waves within the blob. At low $pID$ values this momentum is particularly strong, causing the blob to overshoot the corners after the momentum of oscillatory waves has been established during straighter sections of the course. Under the strongest momentum condition the blob 'crashes' into the circular obstacle at the lower-right of the arena and the blob has to re-form before its progress can continue. An indication of the strength of the momentum can be seen at the end of each course in Fig. 2 where the blob continues to receive attractant input from the final position on the path. Although the position of this stimulus is static, the path of the blob (blue, online) shows that the blob continues to traverse around the periphery of the final position. At low $pID$ (high momentum) values, the radius of this circular movement is much larger than at high $pID$ (low momentum) values.

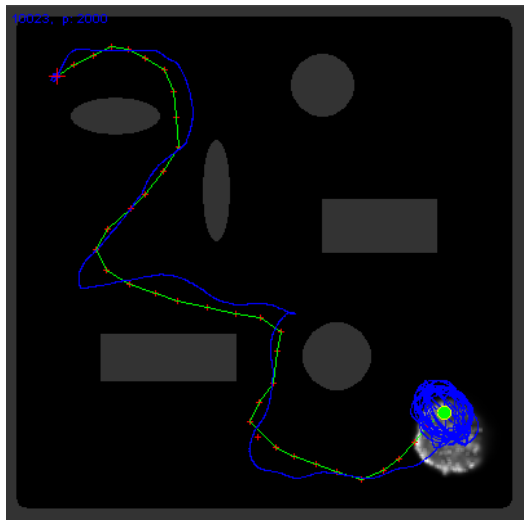

(a)$pID$ 0.001

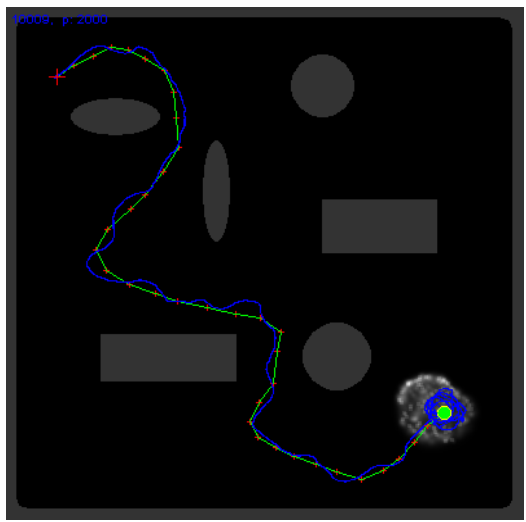

(b)$pID$ 0.05

Fig. 2 Automated guidance of amorphous amoeboid robot via attractants. a) Image showing trajectory of blob (blue online) at $pID$ 0.001 as it is guided along path (green online) from start (cross) to finish (circle), b) Trajectory of blob at $pID$ 0.05.

As the blob traverses the points on the path, there is a repetitive sequence of error minimisation which occurs, as shown in Fig. 3 which shows the distance error at of the arena traversal at $pID$ 0.05 (from Fig. 2b) from start to finish (Fig. 3a) and an enlarged portion of the migration plot of Fig. 3b (between 2000-3000 steps). The guidance mechanism exhibits a characteristic 'sawtooth' profile. As the blob moves forward (attracted by the stimulus point presented at the next path node) the distance between the current centroid of the blob and the target node is minimised (the diminishing diagonal lines of the plot). When this distance is less than the distance between the blob and the *current* node, the new node is selected as the target node. The selection of the next node changes the stimulus point location and also causes a sudden jump in migration error (the vertical lines in the plot). This pattern of minimisation and new target selection occurs until the end node of the path is reached, at which point the blob will circle the final node on the path.

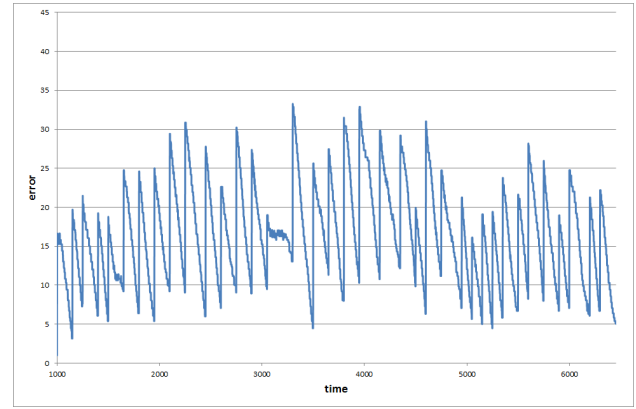

(a)$pID$ 0.05

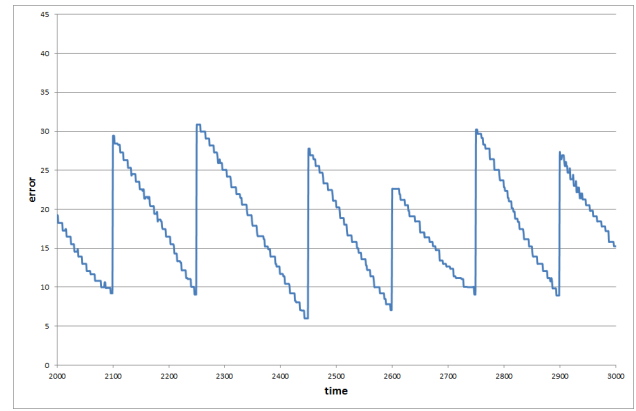

(b)$pID$ 0.05 2000-3000 steps

Fig. 3 Evolution of guidance error during path traversal. a) plot of error distance over time during path traversal at $pID$ 0.05, b) enlarged region of (a) at time 2000-3000 steps. Sawtooth profile plot shows changes in migration error as the blob is guided from node to node along the path, minimising the distance error at the current node (downward diagonal), before the next node is selected, generating a new error distance (vertical line).

### 4.3. Automatic Guidance with Repellent Stimuli

As an alternative to pulling the blob towards the stimulus, it is also possible to 'push' the blob. This can be achieved by mimicking the response of slime mould to hazardous stimuli, for example exposure to light irradiation. In the face of such stimuli slime mould withdraws parts of its body plan away from exposed regions

[22] and can thus be guided away from simple obstacles comprised of light-exposed areas [1]. In the multi-agent model we can reproduce the effect of exposure to light by altering the sensitivity of particles exposed to illuminated regions of the lattice. This reduces flux within that region of the blob. Due to the innate cohesion of the blob, the travelling waves moving within the blob are stronger in unilluminated regions and this attracts particles in exposed regions, propelling the blob away from the stimuli.

We can implement automated guidance by repellent stimuli by having the stimulus point represent an *absence* of illumination, for example a square masked region. *Outside* this region all other areas are temporarily exposed to simulated light exposure. This tends to maintain the blob within the confines of the masked region and also move peripheral parts of the blob that are outside of the protective masked region back inside the mask.

We tested the repellent method on the same obstacle arena as used in the attractant stimulus condition, with the same population size and sensory parameter settings. Fig. 4 shows examples of blob guidance through this arena at different $pID$ values. The general 'sawtooth' pattern of error minimisation along the path is the same as in the attractant guidance method. However, the trajectory of the blob under light irradiation guidance shows much closer adherence to the original path and there is significantly less 'overshoot' than in the attractant guidance method when sudden changes in path direction occur. Across multiple runs of both attractant and repellent conditions the time taken to traverse the path was also shorter in the light-irradiation condition at all $pID$ values (Fig. 5a). Furthermore, the mean error from the desired path was also lower for the light-irradiation condition, compared to the attractant guided method (Fig. 5b).

Why does the light irradiation guidance method track the path more accurately than the attractant method? The square masked region surrounding the blob (for example, Fig. 4a) illuminates all regions outside the mask (i.e. outside of the main blob region), providing simultaneous stimuli at different parts of the blob, compared to the single point stimulus in the attractant condition. Any particles in the illuminated region are subject to the reduction in flux and thus try to return to the unexposed region within the square. This suppression of flux outside the masked square also has the effect of damping travelling waves outside the square, causing less momentum to build up, and more accurate traversal of the path.

It should be noted that although the high-momentum travelling waves in the attractant condition were responsible for the increased error from the desired path, the oscillatory travelling waves are of critical importance for the blob movement. Indeed in the attractant stimulus method, the migration of the blob along the entire course could not be completed without oscillatory movement. In the light irradiation stimulus condition, the light mask was sufficient to move the non-oscillatory collective along the path, but the penalty was a greatly increased time of traversal — over 80,000 scheduler steps — compared to a range of 1900-8000 steps under oscillatory conditions.

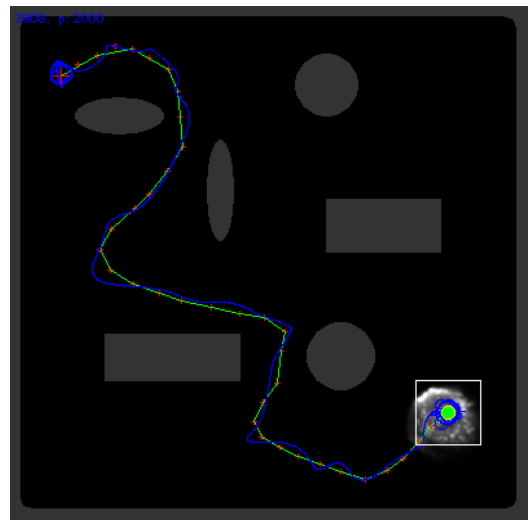

(a) $pID$ 0.001

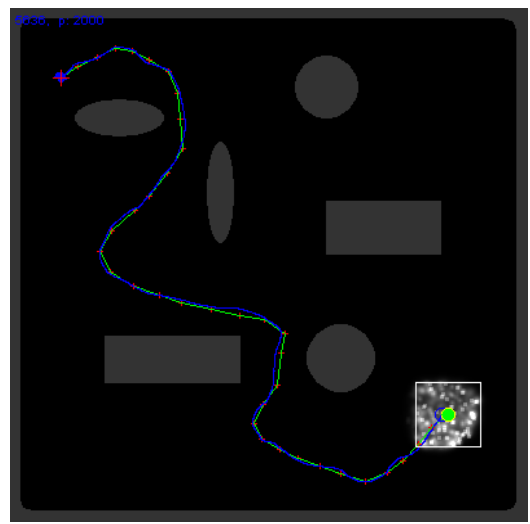

(b) $pID$ 0.05

Fig. 4 Automated guidance of amorphous amoeboid robot via repellent stimulus of simulated light irradiation. a) Image showing trajectory of blob (blue online) at $pID$ 0.001 as it is guided along path (green online) from start (cross) to finish (circle). Light irradiation masked area is indicated by the square region surrounding the blob, b) Example of blob path trajectory at $pID$ 0.05.

### 4.4. Novel Properties of Guided Amoeboid Movement

In addition to the tracking abilities of the hybrid unconventional/conventional computing guidance methods, the amorphous and adaptive properties of the collective result in some interesting properties during its movement. Fig. 6 shows the guidance of the blob along a vertical arena (in this example by repellent light irradiation stimuli). The arena is composed of a narrow channel, some horizontal blocks and finally a very narrow grating, before the destination site. As the blob passes through the narrow channel, the blob elongates, adapting its shape automatically in order to fit through the narrow chan-nel (Fig. 7a) before restoring its approximately circular shape once the channel has been passed.

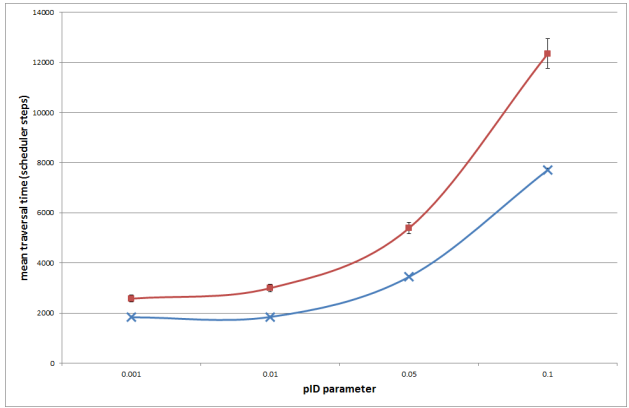

(a) path traversal time

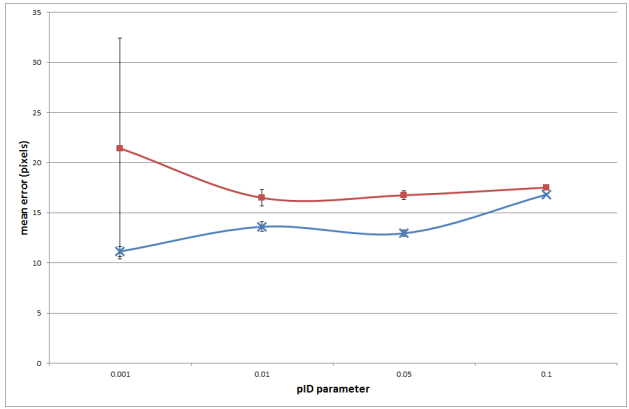

(b) guidance errors

Fig. 5 Comparison of attractant and repellent guidance methods in terms of time and guidance errors. a) comparison of path traversal time for different $pID$ values for attractant (squares) and repellent (crosses) stimuli, b) mean guidance error from path at different $pID$ values for attractant (squares) and repellent (crosses) stimuli (mean of ten runs per $pID$ value, standard deviation indicated). Guidance by repellent stimuli is faster and more accurate.

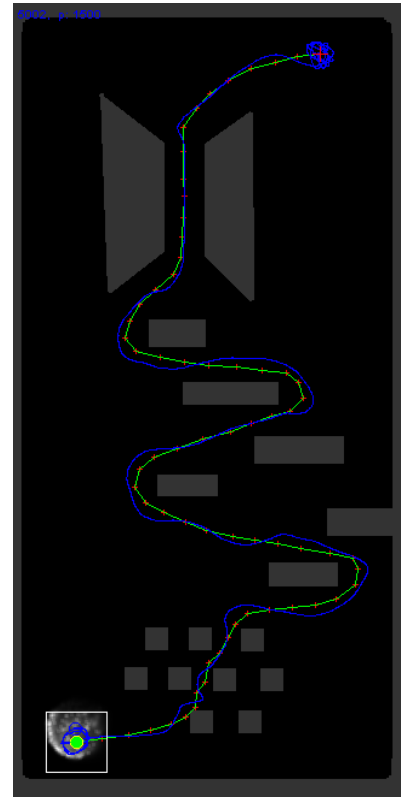

Fig. 6 Traversal of an arena with narrow paths, obstacles and grating by repellent stimuli at $pID$ 0.001. Target path (green, online), and blob trajectory (blue, online) are indicated.

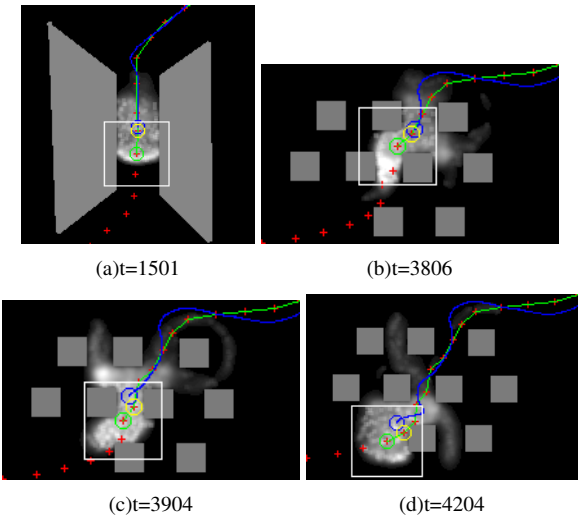

(a) t=1501  (b) t=3806

(c) t=3904  (d) t=4204

Fig. 7 Novel properties of the amoeboid blob during movement. a) blob elongates as it passes through a narrow tunnel, b-d) blob is distorted as it is forced through a narrow grating by the stimulus mask before re-forming its shape when the obstacle is passed.

In the case of the grating at the bottom of the arena, the space between the grating obstacles is so narrow that the blob shape deforms dramatically in regions outside the mask, forming writhing pseudopodium-like tendrils (Fig. 7b-c). Again, once the grating has been crossed the blob reforms its shape as it moves to the goal site. Despite this significant distortion of blob shape the path taken by the blob is fairly close to the target path (Fig. 7d). These properties are a function of the unconventional computing part of the system, in that they are an emergent function of the low-level particle interactions.

The distortion and re-formation of the blob shape at the narrow grating does not occur reliably with the attractant based guidance method, however. Fig. 8 shows a blob guided by a single attractant source entering the grating region (Fig. 8a) where its body plan is distorted on contact with the obstacles (Fig. 8b). The blob becomes entwined on a single obstacle in the grating (Fig. 8c) and minimises its shape to wrap around the obstacle (Fig. 8d). The blob remains in this position indefinitely, cycling around this obstacle. Corruption of the $X,Y$ stimulus location values with Gaussian noise (to try to present multiple stimulus sites) does not detach the blob from the obstacle. This behaviour again demonstrates the effectiveness of guidance by illumination masked regions compared to the attractant guidance method. Why does the blob not become stuck at this obstacle when guided by the repellent mask? Again, this is because the illumination mask presents multiple stimulus points to the blob (at the interface of the square mask edges which contact the blob), whereas the attractant guidance method only presents a single guidance stimulus.

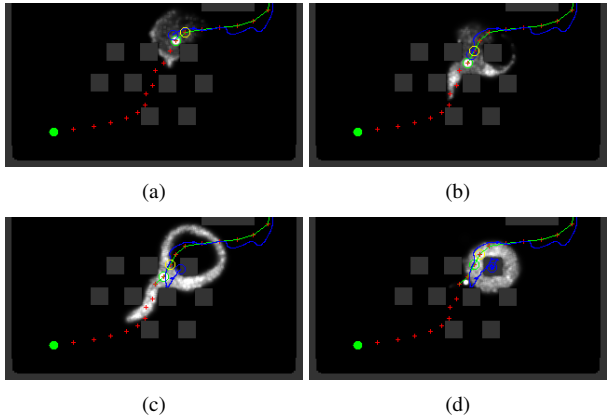

Fig. 8 Guided blob becomes trapped on grating using attractant method. a) blob enters grating area, b) distortion of blob pattern occurs, c) blob surrounds grating obstacle, d) blob minimises shape around grating, continuing to cycle in this position.

## 5. CONCLUSIONS

In this report we have examined the automated control and guidance of collective movement in an unconventional computing substrate, a multi-agent system inspired by slime mould *Physarum polycephalum*. Collective movement was generated in a morphologically adaptive 'blob' comprising a population of simple particles on a diffusive lattice. Taking inspiration from slime mould, the position of this blob could be altered by the spatial placement of attractant and repellent stimuli. The self-organised generation of blob cohesion and its movement by oscillatory travelling waves was combined with a closed-loop feedback mechanism to calculate the current position of the blob in relation to a pre-defined path. By comparing the current blob position with the closest point on the path, a stimulus (the next available path point) was then presented to the unconventional computing substrate, causing the blob to migrate along the path. Of the two stimulus types investigated (attractant and repellent) to guide the blob, the repellent stimulus (masking the blob from simulated light irradiation) resulted in a faster path traversal with fewer errors (in terms of distance from the pre-defined path) and allowed the blob to pass automatically through very narrow gratings. Passage through a narrow grating could not be achieved using the attractant stimulus condition, due to the lack of simultaneous stimuli to the blob when compared to the repellent mask method. The momentum of the blob could be controlled by adjusting a parameter of the model. Stronger momentum resulted in faster path traversal in both stimulus types, but resulted in a characteristic overshooting of path corners in the attractant stimulus condition. We hope that this work will provide a contribution towards future implementations of control and guidance in soft-bodied robotics which exploit collective movement in swarm systems.

## ACKNOWLEDGEMENTS

This research was supported by the EU research project "Physarum Chip: Growing Computers from Slime Mould" (FP7 ICT Ref 316366).